\documentclass[pre,showpacs,preprint]{revtex4}

\usepackage{amsmath}
\usepackage{graphicx}
\usepackage{dcolumn}
\usepackage[mathscr]{euscript}
\usepackage{bm}
 
\DeclareMathOperator\erfi{erfi}

\DeclareMathOperator{\sgn}{sgn}

\begin{document}

\title{  Inhomogeneous cooling state of a strongly confined granular gas at low density }
\author{J. Javier Brey, M.I. Garc\'{\i}a de Soria, and P. Maynar}
\affiliation{F\'{\i}sica Te\'{o}rica, Universidad de Sevilla,
Apartado de Correos 1065, E-41080, Sevilla, Spain}
\date{\today }

\begin{abstract}
The inhomogeneous cooling state describing the hydrodynamic behavior of a  freely evolving granular gas strongly confined between two parallel plates is studied, using a Boltzmann kinetic equation derived recently. By extending the idea of the homogeneous cooling state, we propose a scaling distribution in which all the time dependence occurs through the granular temperature of the system, while there is a dependence on the distance to the confining walls through the density. It is obtained that the velocity distribution is not isotropic, and it has two different granular temperature parameters associated to the motion perpendicular and parallel to the confining plates, respectively, although their cooling rates are the same. Moreover, when approaching the inhomogeneous cooling state,  energy is sometimes transferred from degrees of freedom with lower granular temperature to those with a  higher one, contrary to what happens in molecular systems. The cooling rate and the two partial granular temperatures are calculated by means of a Gaussian approximation. The theoretical predictions are compared with molecular dynamics simulation results and a good agreement is found.

\end{abstract}

\maketitle

\section{Introduction}
\label{s1}
Granular gases are systems composed of macroscopic particles which do not conserve kinetic energy when they collide \cite{Go03}. As a consequence of the energy dissipation, they are intrinsically  non-equilibrium systems and have a  very rich and peculiar phenomenology \cite{JNyB96}, so they can be considered as a proving ground for non-equilibrium statistical mechanics. In particular, a monolayer of spherical macroscopic particles is perhaps the simplest experimental system exhibiting a variety of phenomena, including  non-Maxwellian velocity distributions, phase transitions with a rather complex phase diagram, long lived fluctuations,  and other non-trivial non-equilibrium effects  \cite{OyU99,Metal05,RPGRSCyM11,CMyS12, ASOyU08}. Most of these behaviors look quite similar to other observed in normal, molecular fluids, and that are successfully described by hydrodynamics. The analysis of granular gases via kinetic theory has been an active field of research in the last two decades, and it has been shown that many features of granular systems can be explained by means of a fluid of hard spheres with smooth  inelastic collisions. An important objective of the studies has been the derivation of hydrodynamic-like equations, with explicit expressions for the corresponding transport coefficients. Once a given kinetic equation has been formulated (e.g extensions of  the Boltzmann equation for a low density gas or of the revised Enskog equation for hard spheres at higher density), the usual Chapman-Enskog method has been adapted to the case of inelastic collisions \cite{BDKyS98,GyD99a}. The key ingredient of the method is the search of a ``normal''  solution of the kinetic equation, i.e. a solution in which all the space and time dependence occurs through the hydrodynamic fields characterizing the macroscopic description of the fluid. In practice, this requires to consider an expansion around some reference state. For elastic collisions, this state is the local equilibrium Maxwellian, while for inelastic collisions the reference state is the local homogenous cooling state (HCS), with a time-dependent granular temperature, consequence of the energy loss in collisions. Moreover, the velocity distribution  of the HCS deviates significantly from  a Maxwellian \cite{GyS95,BRyC96,vNyE98}.

The above research program has been carried out up to now for bulk systems, i.e. focusing in regions far away from the boundaries of the system. The effect of the physical boundaries  occurs only through the boundary conditions for the kinetic equations and also for the macroscopic, hydrodynamic equations.  The objective here is the search of a similar reference state  for the case of a strongly confined granular gas with slit geometry. When the system is under extreme confinement, this constraint shows up not only through the boundary conditions, but also affects the form of the kinetic equation itself and hence of the macroscopic evolution equations.

The starting point of the present analysis will be the extension to inelastic particles of a Boltzmann-like kinetic equation formulated recently for a system of elastic hard spheres confined between two infinite parallel plates at rest \cite{BGyM17,BMyG16}. For this case, it was shown that the equilibrium velocity distribution is a Maxwellian, with a uniform temperature, while  there is  a nonuniform density profile along the direction perpendicular to the plates confining the system. The modifications introduced by the inelasticity  produce relevant qualitative differences when comparing with the elastic case. In the context of hydrodynamics and normal solution, the distribution function of the inelastic confined system is defined such that all the time dependence occurs through the local granular temperature of the system, while the density depends on the distance  to the walls.  For this reason, this state will be referred to as the inhomogeneous cooling state (ICS) of the strongly confined system. If one considers the partial granular temperature parameters associated to the motion parallel and perpendicular to the plates,  they are different. The combination of strong confinement and inelasticity renders the velocity distribution of the system anisotropic. When valuing this feature, it must be kept in mind that no external energy is being injected into the system when it is in the ICS, but it is freely evolving. Consequently, the symmetry breaking in velocity space is induced by the inelasticity of the  collisions between particles combined with the confinement.

The remaining of this paper is organized as follows. In the next section the Boltzmann equation for a strongly confined gas of hard spheres \cite{BGyM17,BMyG16} is shortly reviewed and extended to the case of smooth inelastic particles. Also, the inhomogeneous cooling state (ICS) is introduced and the condition of isotropic cooling rate is deduced. In Sec. \ref{s3}, it is shown that the density profile is not uniform along the direction perpendicular to the plates, obtaining an expression for it. This is similar to what happens in the equilibrium state of a strongly confined elastic gas, but the shape of the profile depends on the inelasticity of collisions. Approximate expressions for the cooling rates in the directions parallel and perpendicular to the plates are obtained in Sec. \ref{s4} and, from them, an expression for the constant granular  temperature ratio in the ICS follows. The result is expected to apply  for not too strong inelasticity and for a separation of the two plates close to the diameter of the hard spheres. The theoretical predictions are compared with molecular dynamics simulation data in Sec. \ref{s5}. A good agreement is found for both, the density profile and the granular temperature ratio. The final section of the paper contains a short summary and also a discussion of the possibility that the energy goes from the ``cooler'' degrees of freedom to the ``hotter'' ones in the process of approaching the ICS. Some details of the calculations are presented in the Appendices.

\section{Boltzmann equation and inhomogeneous cooling state}
\label{s2}
The system we consider is an ensemble of $N$ smooth  inelastic hard spheres of mass $m$ and diameter $\sigma$, confined between two horizontal parallel hard plates separated a distance $h$, smaller than two particle diameters, $\sigma < h < 2 \sigma$. Inelasticity of collisions between particles is characterized by a constant, velocity-independent, coefficient of normal restitution $\alpha$, in the range $ 0 < \alpha \leq 1$.  It is assumed that in the low-density limit, the one-particle distribution function, $f({\bm r},{\bm v},t)$, for the density of particles at position ${\bm r}$, with velocity ${\bm v}$ at time $t$, is well described by the Boltzmann-like kinetic equation
\begin{equation}
\label{2.1}
\frac{\partial f}{\partial t}+{\bm v} \cdot \frac{\partial f}{\partial {\bm r}}= J[{\bm r},{\bm v}|f].
\end{equation}
The collision term $J[{\bm r},{\bm v}|f]$ for the scattering of two particles is
\begin{eqnarray}
\label{2.2} 
J[{\bm r},{\bm v}|f] & \equiv & \sigma  \int d{\bm v}_{1} \int_{0}^{2\pi} d \varphi \int_{\sigma/2}^{h-\sigma/2} dz_{1}\,   | {\bm g} \cdot \widehat {\bm \sigma} | \nonumber \\
&& \times  \left[ \Theta ({\bm g} \cdot \widehat {\bm \sigma})  \alpha^{-2} b_{\bm \sigma}^{-1}- \Theta (-{\bm g} \cdot \widehat{\bm \sigma})\right]f({\bm r}_{1},{\bm v}_{1} ,t) f({\bm r},{\bm v},t),
\end{eqnarray}
where ${\bm r} \equiv \{x,y,z \}$, ${\bm r}_{1} \equiv \{x,y,z_{1} \}$, ${\bm g} \equiv {\bm v}_{1}- {\bm v}$, and $\Theta$ is the Heaviside step function. The $z$ axis has been taken perpendicular to the hard plates, with its origin located at one of them, and the positive direction pointing inside the system. Moreover,  $ \widehat{\bm \sigma}$ is a unit vector along the line of the two colliding particles at contact. With the coordinate system we are using, in which $\theta$ and $\varphi$ are the polar and azimuthal angles, respectively,  $\widehat{\bm \sigma} =\{ \sin \theta \sin \varphi, \sin \theta \cos \varphi, \cos \theta \}$, with $\cos \theta = (z_{1}-z)/\sigma$. Finally, the operator $b_{\bm \sigma}^{-1}$ changes all the velocities ${\bm v}$ and ${\bm v}_{1}$ to its right into their pre-collisional values ${\bm v}^{*}$ and ${\bm v}^{*}_{1}$, respectively. For a system with the same geometry, but composed of elastic hard spheres, the Boltzmann equation has been derived using  arguments similar to those employed to obtain the usual Boltzmann equation for unconfined systems \cite{BGyM17}. The arguments can  can be directly extended to the present case of inelastic hard spheres \cite{BDyS97}. The only and fundamental difference is that for smooth inelastic collisions the pre-collisional velocities are given by
\begin{equation} 
\label{2.3}
{\bm v}^{*} \equiv b_{\bm \sigma}^{-1} {\bm v} = {\bm v} + \frac{1 + \alpha}{2 \alpha} \left( {\bm g} \cdot \widehat{\bm \sigma} \right) \widehat{\bm \sigma},
\end{equation}
\begin{equation} 
\label{2.4}
{\bm v}^{*}_{1} \equiv b_{\bm \sigma}^{-1} {\bm v}_{1} = {\bm v}_{1}-  \frac{1 + \alpha}{2 \alpha} \left( {\bm g} \cdot \widehat{\bm \sigma} \right) \widehat{\bm \sigma}.
\end{equation}
The kinetic equation (\ref{2.1}) has to be solved with the appropriated boundary conditions, e.g. elastic walls, thermal walls, and so on. A useful identity, for an arbitrary function $\chi({\bm v})$ is
\begin{eqnarray}
\label{2.5}
\int d{\bm v }\,  \chi ({\bm v}) J[{\bm r},{\bm v} |f] &=& \sigma \int d{\bm v} \int d {\bm v}_{1}  \int_{0}^{2\pi} d \varphi \int_{\sigma/2}^{h-\sigma/2} dz_{1}\,   | {\bm g} \cdot \widehat {\bm \sigma} | \Theta (-{\bm g} \cdot \widehat{\bm \sigma}) \nonumber \\
&& \times  f({\bm r}_{1},{\bm v}_{1} ,t) f({\bm r},{\bm v},t) \left( b_{\bm \sigma} -1 \right) \chi ({\bm v}),
\end{eqnarray}
where $b_{\bm \sigma}$ is the operator for direct collisions,
\begin{equation}
\label{2.6}
{\bm v}^{\prime} \equiv b_{\bm \sigma} {\bm v} = {\bm v} + \frac{1 + \alpha}{2} \left( {\bm g} \cdot \widehat{\bm \sigma} \right) \widehat{\bm \sigma},
\end{equation}
\begin{equation} 
\label{2.7}
{\bm v}^{\prime}_{1} \equiv b_{\bm \sigma} {\bm v}_{1} = {\bm v}_{1} - \frac{1 + \alpha}{2 } \left( {\bm g} \cdot \widehat{\bm \sigma} \right) \widehat{\bm \sigma}.
\end{equation}
Equation (\ref{2.5}) implies
\begin{eqnarray}
\label{2.8}
\int_{\sigma/2}^{h-\sigma/2} dz \int d{\bm v }\,  \chi({\bm v}) J[{\bm r},{\bm v} |f] &=& \frac{\sigma}{2} \int d{\bm v} \int d {\bm v}_{1}  \int_{0}^{2\pi} d \varphi 
\int_{\sigma/2}^{h-\sigma/2} dz \int_{\sigma/2}^{h-\sigma/2} dz_{1}\,   | {\bm g} \cdot \widehat {\bm \sigma} | \Theta (-{\bm g} \cdot \widehat{\bm \sigma}) \nonumber \\
&& \times  f({\bm r}_{1},{\bm v}_{1} ,t) f({\bm r},{\bm v},t) \left( b_{\bm \sigma} -1 \right) \left[ \chi({\bm v})+\chi({\bm v}_{1}) \right].
\end{eqnarray}
Macroscopic fields are introduced in the usual way: local number density, $n({\bm r},t)$, local velocity flow, ${\bm u}({\bm r},t)$, and local granular temperature, $T({\bm r},t)$, are defined by
\begin{equation}
\label{2.9}
n({\bm r},t) \equiv \int d{\bm v}\, f({\bm r}, {\bm v},t),
\end{equation}
\begin{equation}
\label{2.10}
n({\bm r},t) {\bm u}({\bm r},t) \equiv  \int d{\bm v}\, {\bm v} f({\bm r}, {\bm v},t),
\end{equation}
\begin{equation}
\label{2.11}
\frac{3}{2} n({\bm r},t) T({\bm r},t)  \equiv \frac{1}{2}  \int d{\bm v}\, m \left[ {\bm v}-{\bm u}({\bm r},t) \right]^{2}f({\bm r}, {\bm v},t),
\end{equation}
respectively. In the following, it will be convenient to consider also partial granular temperatures associated to the $z$ direction, $T_{z}({\bm r},t)$,  and to the velocity vector parallel to the plates, i.e. perpendicular to the $z$ axis, $T_{=}({\bm r},t)$. They are defined as
\begin{equation}
\label{2.12}
 n({\bm r},t) T_{z}({\bm r},t) \equiv \int d{\bm v}\, m \left[ v_{z}-u_{z}({\bm r},t) \right]^{2}f({\bm r}, {\bm v},t),
 \end{equation}
 \begin{equation}
 \label{2.13}
 n({\bm r},t) T_{=}({\bm r},t) \equiv \frac{1}{2}  \int d{\bm v}\, m \left\{ \left[ (v_{x}-u_{x}({\bm r},t) \right]^{2} +\left[ (v_{y}-u_{y}({\bm r},t) \right]^{2} \right\}  f({\bm r}, {\bm v},t).
 \end{equation}
 It is
 \begin{equation}
 \label{2.14}
 T({\bm r},t) = \frac{T_{z}({\bm r},t)+ 2T_{=}({\bm r},t)}{3}\, .
 \end{equation}
 On the basis of the studies of bulk systems of inelastic hard spheres \cite{GyS95,BRyC96,vNyE98} and of the results for  quasi-two-dimensional systems of elastic spheres \cite{BMyG16,BBGyM17}, it is assumed that there is a special normal solution of the kinetic equation for which all the time dependence of the distribution function occurs through the granular temperature and all the space dependence takes place through the dependence of the number density on the $z$ coordinate. Moreover, there is no macroscopic velocity flow. More specifically, it is assumed that there are solutions of the form
 \begin{equation}
 \label{2.15}
 f_{0}(z,{\bm v},t)= n(z) v_{0}^{-3}(t) \phi({\bm c}_{=},c_{z}),
 \end{equation} 
where 
\begin{equation}
\label{2.16}
v_{0}(t) \equiv \left[ \frac{2T(t)}{m} \right]^{1/2}
\end{equation}
is a local thermal velocity defined in terms of the temperature $T(t)$,
\begin{equation}
\label{2.17}
{\bm c} \equiv \frac{\bm v}{v_{0}(t)}\, ,
\end{equation}
and ${\bm c}_{=} $ is the two-dimensional vector defined by the components of ${\bm v}$   in the plane parallel to the hard walls. It is worth to stress that the distribution function is not assumed to be isotropic, although by symmetry considerations it does not depend on the direction of ${\bm c}_{=}$ or the sign of $c_{z}$, i.e. it is a function of $|{\bm c}_{=}|$ and $|c_{z}|$. For the same reason,  the density field is assumed to be symmetric around the plane $z=h/2$, i.e.
\begin{equation}
\label{2.18}
n(z)=n(h-z),
\end{equation}
$\sigma/2 < z < h- \sigma/2$. The above state can be viewed as an extension of the homogeneous cooling state  of  freely evolving bulk granular systems \cite{GyS95,BRyC96,vNyE98} to strongly confined ones, and it will be referred to as the inhomogeneous cooling state (ICS). Its distribution function verifies the boundary conditions corresponding to elastic walls at rest, namely \cite{BGyM17}
\begin{equation}
\label{2.19}
\Theta (v_{z}) f_{0}(z,{\bm v}, t) \delta \left( z- \frac{\sigma}{2} \right) = \Theta (v_{z}) f_{0}(z,{\bm v}^{(w)},t) \delta \left ( z- \frac{\sigma}{2} \right),
\end{equation}
\begin{equation}
\label{2.20}
 \Theta (-v_{z}) f_{0}(z,{\bm v}, t) \delta \left( z- h+\frac{\sigma}{2} \right) = \Theta (-v_{z}) f_{0}(z,{\bm v}^{(w)},t) \delta \left ( z- h+\frac{\sigma}{2} \right),
\end{equation} 
with 
\begin{equation}
\label{2.21}
{\bm v}^{(w)}= {\bm v}- 2 v_{z} \widehat{\bm e}_{z},
\end{equation}
with $ \widehat{\bm e}_{z}$ being the unit vector in the positive direction of the $z$ axis. These relations express the conservation of the flux of particles at the walls. The dimensionless velocity distribution function $\phi$  in Eq. (\ref{2.15}) must verify the conditions
\begin{equation}
\label{2.22}
\int d{\bm c}\,  \phi ({\bm c}_{=},c_{z})=1,
\end{equation}
\begin{equation}
\label{2.23}
\int d{\bm c}\,  {\bm c} \phi ({\bm c}_{=},c_{z})=0,
\end{equation}
\begin{equation}
\label{2.24}
\int d{\bm c}\,  c^{2} \phi ({\bm c}_{=},c_{z})=\frac{3}{2},
\end{equation}
that follow directly from the definition of the macroscopic fields, Eqs.\ (\ref{2.9})-(\ref{2.11}). In terms of $\phi$, the partial granular temperatures are given by
\begin{equation}
\label{2.25}
T_{z}(t)= 2 T(t) \int d{\bm c}\,  c_{z}^{2} \phi ({\bm c}_{=}, c_{z}),
\end{equation}
\begin{equation}
\label{2.26}
T_{=}(t)= T(t) \int d{\bm c}\, c_{=}^{2} \phi ({\bm c}_{=}, c_{z}).
\end{equation}
It follows that the two partial temperatures are proportional to each other and their ratio is a constant, independent from  time $t$,
\begin{equation}
\label{2.27}
\frac{T_{z}(t)}{T_{=}(t)} = \gamma = \mbox{const}.
\end{equation}
One possibility is that the two partial temperatures, and the global one, are equal, as it is the case for a system of elastic hard spheres strongly confined at equilibrium \cite{BMyG16,BGyM17}. However, this can not be assumed {\em a priori} and the proportionality constant $\gamma$ must be determined from the solution of the kinetic equation. Actually, it will be found below that, if the inhomogeneous cooling state  as defined by Eq.\ (\ref{2.15}) exits,  the two partial temperatures, $T_{z}$ and $T_{=}$, must be  different. A similar behavior is found in the homogeneous cooling state of a granular mixture for the partial temperatures of the species \cite{GyD99}, being responsible of relevant macroscopic effects, as segregation \cite{BRyM05}. From Eqs.\ (\ref{2.14}) and (\ref{2.27}) one gets
\begin{equation}
\label{2.28}
T_{z}(t)= \frac{3\gamma}{\gamma +2}\,  T(t),
\quad T_{=}(t)= \frac{3}{\gamma +2}\, T(t).
\end{equation}
 Evolution equations for the two partial granular temperatures in the ICS are derived from the Boltzmann equation (\ref{2.1}), by using the expression for the distribution function given in Eq. (\ref{2.15}),
 \begin{equation}
 \label{2.29}
 \frac{\partial T_{z} (t)}{\partial t} = - \zeta_{z}T_{z}(t), \quad \quad \frac{\partial T_{=}(t)}{\partial t} = - \zeta_{=}T_{=}(t).
 \end{equation}
 The cooling rates (fractional energy changes per unit of time) have the form
 \begin{equation}
 \label{2.30}
 \zeta_{z}= \frac{v_{0}(t)}{\sigma}\, \zeta^{*}_{z}, \quad \quad \zeta_{=}= \frac{v_{0}(t)}{\sigma}\, \zeta^{*}_{=},
 \end{equation}
 with the dimensionless cooling rates given by
 \begin{equation}
 \label{2.31}
 \zeta^{*}_{z} = -\frac{2(\gamma +2)}{3 \gamma } \int d{\bm c}\,  c_{z}^{2} J^{*} [z^{*},{\bm c}|\phi],
 \end{equation}
 \begin{equation}
 \label{2.32}
 \zeta^{*}_{=}= -\frac{\gamma +2}{3} \int d{\bm c}\,  c_{=}^{2} J^{*} [z^{*},{\bm c}|\phi].
 \end{equation}
In the above expressions, $J^{*}[z^{*},{\bm c}|\phi]$ is the dimensionless collision term
\begin{eqnarray}
\label{2.33}
J^{*} [z^{*},{\bm c}| \phi] &=& \int d{\bm c}_{1}  \int_{0}^{2\pi} d \varphi\  \int_{1/2}^{h/ \sigma-1/2} d z_{1}^{*}\, |{\bm c}_{10} \cdot \widehat{\bm \sigma}| n^{*}(z^{*}_{1})
 \nonumber \\
 && \left[ \Theta \left( {\bm c}_{10} \cdot \widehat{\bm \sigma} \right) \alpha^{-2} b_{\bm \sigma}^{-1} - \Theta \left( {-\bm c}_{10} \cdot \widehat{\bm \sigma} \right) \right] \phi({\bm c}_{1 =},c_{1z}) \phi ({\bm c}_{=},c_{z}),
 \end{eqnarray}
 where ${\bm c}_{10} \equiv {\bm c}_{1}-{\bm c}$ and
 \begin{equation}
 \label{2.34}
 z^{*} \equiv  z/\sigma, \quad \quad n^{*}(z^{*})  \equiv n(z) \sigma^{3}.
 \end{equation}
 Of course, the action of the operator $b_{\bm \sigma}^{-1}$ over a function of ${\bm c}$ and ${\bm c}_{1}$ is still given by Eqs. (\ref{2.3}) and (\ref{2.4}), but replacing the original velocities ${\bm v}$ by the scaled ones ${\bm c}$.
 
 Note that, as a consequence of the form of the distribution function of the ICS, the dimensionless cooling rates turn out to be constant, independent of both $z$ and $t$. Time derivative of Eq. (\ref{2.27}) leads to the relevant consequence that the cooling rates defined by Eqs.\ (\ref{2.29}) are equal. 
 \begin{equation}
 \label{2.35}
 \zeta_{z}(t) = \zeta_{=}(t)= \zeta (t).
 \end{equation}
Moreover, from (\ref{2.14}),
\begin{equation}
\label{2.36}
\frac{\partial T(t)}{\partial t}=- \zeta(t) T(t).
\end{equation}
To identify the function $\phi({\bm c}_{=}, c_{z})$, Eq. (\ref{2.15}) is substituted into the kinetic equation (\ref{2.1}), and reduced variables $z^{*}$ and ${\bm c}$ are used. Then the equation becomes
\begin{equation}
\label{2.37}
\frac{\zeta^{*}}{2} \frac{\partial}{\partial  {\bm c}}\,\cdot \left[{\bm c} \phi({\bm c}_{=},c_{z}) \right] + \frac{\partial \ln n^{*}(z^{*})}{\partial z^{*}} c_{z} \phi({\bm c}_{=},c_{z})   = J^{*}[z^{*},{\bm c}|\phi].
\end{equation}
with $\zeta^{*} = \sigma \zeta/v_{0}(t)$. In summary, the distribution function of the assumed ICS is given by the solution of the dimensionless kinetic equation (\ref{2.37}), with the cooling rate given by either Eq. (\ref{2.31}) or Eq. (\ref{2.32}), and the value of $\gamma$ being determined by Eq.(\ref{2.35}) or, equivalently,
\begin{equation}
\label{2.38}
\gamma \equiv \frac{T_{z}(t)}{T_{=}(t)} = \frac{2 \int d{\bm c}\,  c_{z}^{2} J^{*}[z^{*},{\bm c}|\phi]}{\int d{\bm c}\,   c_{=}^{2} J^{*}[z^{*},{\bm c}|\phi]}\, .
\end{equation}
Consistency requires that $\gamma$, as given by this expression, be time and position-independent. This is a strong theoretical prediction, following from the existence itself of the ICS. Both equations, (\ref{2.37}) and (\ref{2.38}), must be solved self-consistently for  the function $\phi({\bm c}_{=}, c_{z})$ and the temperature ratio $\gamma$.

\section{Density profile}
\label{s3}

 In this section, an expression for the density profile $n(z)$ of the ICS will be obtained.  By construction, $n(z)$ does not depend on time, but it happens to depend on the temperature ratio $\gamma$, that must be determined self-consistently, as discussed in the previous section. Velocity integration of the kinetic equation (\ref{2.37}) does not provide any relevant physical information, since all the terms vanish identically. On the other hand, multiplication of the equation by $c_{z}$ and latter integration over ${\bm c}$ gives
 \begin{equation}
 \label{3.1}
 \frac{\partial \ln n^{*}(z^{*})}{\partial z^{*}}= \frac{\pi (1+\alpha)}{\gamma} \int _{1/2}^{h/\sigma-1/2} d z^{*}_{1}\,  \left[ z^{*}-z^{*}_{1}+ ( z^{*}-z^{*}_{1})^{3} (\gamma-1)\right] n^{*}(z^{*}_{1}).
 \end{equation} 
Details of the calculations are given in Appendix \ref{a}. Multiplication of the kinetic equation by ${\bm c}_{=}$ before velocity integration leads to trivial identities. In the equilibrium elastic limit, $\alpha=\gamma=1$, Eq. (\ref{3.1})  reduces to
\begin{equation}
\label{3.2}
 \frac{\partial \ln n^{*}(z^{*})}{\partial z^{*}}= 2 \pi \int _{1/2}^{h/\sigma-1/2}  dz^{*}_{1}\,    (z^{*}-z^{*}_{1}) n^{*}(z_{1}),
 \end{equation}
that agrees with the result reported in Ref. \cite{BMyG16}.  It is worth to stress that Eq.\  (\ref{3.1}) has been derived without assuming any specific form for the velocity distribution $\phi$, by using only its assumed symmetry properties. On the other hand, the equation is not closed, since it contains the temperature ratio $\gamma$, that has to be obtained from elsewhere. By introducing the variable $\eta_{1}=z^{*}_{1}-h/2\sigma$, and exploiting  the symmetry of the density profile, $n^{*}(\eta_{1})= n^{*} (-\eta_{1})$, Eq. (\ref{3.1}) can be expressed in the equivalent form
\begin{eqnarray}
\label{3.3}
 \frac{\partial \ln n^{*}(z^{*})}{\partial z^{*}} & =  &\frac{\pi (1+\alpha)N \sigma^{2}}{ \gamma A} \left[z^{*}-\frac{h}{2\sigma } + \left( z^{*}-\frac{h}{2\sigma} \right)^{3}
 (\gamma-1) \right] \nonumber \\
&& + \frac{ 3 \pi (1+\alpha)}{ \gamma} \left( z^{*}-\frac{h}{2\sigma} \right) \left( \gamma -1) \right) \int _{-h/2\sigma+1/2}^{h/ 2\sigma -1/2} d \eta_{1}\,  \eta_{1}^{2} n^{*} (\eta_{1}),
\end{eqnarray}
where $A$ is the area of each of the two parallel plates confining the granular gas. This is a linear integro-differential equation,  that can be solved by means of numerical methods. Nevertheless, for the system geometry we are considering, it is
 \begin{equation}
 \label{3.4}
 \left| z^{*}- \frac{h}{2 \sigma} \right| <\frac{h}{2\sigma} -\frac{1}{2} < \frac{1}{2},
 \end{equation}    
 going to zero as $h$ approaches $\sigma$. Moreover, $\gamma$ is expected to be of the order of unity, something that will be confirmed below. As a consequence, a good approximation to Eq. (\ref{3.1}) is expected to be
 \begin{equation}
 \label{3.5}
  \frac{\partial \ln n^{*}(z^{*})}{\partial z^{*}} =  \frac{\pi (1+\alpha)N \sigma^{2}}{ \gamma A} \left(z^{*}-\frac{h}{2\sigma}  \right).
  \end{equation}       
The accuracy will improve as $\alpha$ and $h$ approaches unity and  $\sigma$, respectively.   The solution of this last equation is
\begin{equation}
\label{3.6}
n^{*}(z^{*})= \frac{N\sigma^{2}}{Ab} \exp \left[ a \left( z^{*}- \frac{h}{2 \sigma} \right)^{2} \right].
\end{equation}
with
\begin{equation}
\label{3.7}
a \equiv \frac{\pi (1+\alpha) N \sigma^{2}}{2 \gamma A}\, ,
\end{equation}
\begin{equation}
\label{3.8}
b= \left( \frac{\pi}{a} \right)^{1/2} \erfi \left[ \frac{\sqrt{a}}{2} \left(\frac{h}{\sigma} -1 \right) \right].
\end{equation}
Here $\erfi(y)$ is the imaginary error function defined as 
\begin{equation}
\label{3.9}
\erfi(y) \equiv \pi^{-1/2} \int_{-y}^{y} dy^{\prime}\, e^{y^{\prime2}}.
\end{equation}
The inhomogeneity of the density field in the direction perpendicular to the plates is a direct consequence of the confinement of the system, although the specific shape of the profile, for a given value of the width $h$,  depends on the inelasticity of collisions. Actually, the only change in Eq.\ (\ref{3.6}) when going to the elastic equilibrium limit is in the value of $a$, Eq. (\ref{3.7}), that becomes $a= \pi N \sigma^{2}/A$ in that limit.

\section{Approximate expressions for the cooling rates and the temperature ratio}
\label{s4}

In order to determine the time-dependent temperature fields, the equations for the ICS have to be solved. This is because velocity moments of order larger than two appear in the temperature equations and those moments depend on the form of the velocity distribution. The usual method to solve kinetic equations consists in expanding the distribution function in a complete set of orthogonal polynomials with a Gaussian measure. The coefficients of the expansion are related with the moments of the velocity distribution, and are determined self-consistently  by introducing the polynomial representation into the kinetic equation, multiplying by the appropriated  velocity polynomial, and integrating over the velocity. In practice, a finite, small, number of polynomials is considered. In most of the cases, Sonine polynomials are used, so that the leading term in the expansion is a Gaussian, chosen such that it is normalized to unity and provides the exact value of the second moment, i.e. the granular temperature in our case. Here, the lowest order approximation will be employed to compute the cooling rates. Then, inside the velocity integrals in the right-hand side of Eqs. (\ref{2.31}) and (\ref{2.32}), the dimensionless velocity distribution is approximated by
\begin{equation}
\label{4.1}
\phi({\bm c}_{=},c_{z})  \approx \phi_{=}({\bm c}_{=}) \phi_{z}(c_{z})\, ,
\end{equation}
\begin{equation}
\label{4.2}
\phi_{=}({\bm c}_{=})=  \frac{\gamma +2}{3 \pi}\,  e^{-\frac{\gamma+2}{3} c_{=}^{2}}\, ,
\end{equation}
\begin{equation}
\label{4.3a}
\phi_{z}(c_{z})=  \left( \frac{\gamma +2}{3  \pi \gamma} \right)^{1/2}  e^{-\frac{\gamma+2}{3\gamma} c_{z}^{2}}\, .
\end{equation}
The above estimate involves two different approximations. First, correlations between the $z$-component of the velocity and the velocity components in the parallel  plane are neglected. Second, the marginal velocity distributions of  both, $v_{z}$ and ${\bm v}_{=}$ are approximated by Gaussians. This estimate is suggested by the fact that it is exact in the equilibrium elastic limit \cite{BMyG16}, and it is known to be a very good approximation for a non-confined  system of inelastic rough spheres, where the approximation is made for  the translational and  rotational velocities  \cite{GyS95}. It must be stressed that the approximation introduced refers only to the calculation of velocity integrals giving the cooling rates and not to  the expression of the velocity distributions themselves. From Eq.\ (\ref{2.32}), using the dimensionless form of the property given in Eq. (\ref{2.5}), it is found
\begin{eqnarray}
\label{4.3}
\zeta_{=}^{*} & = &- \frac{\gamma+2}{3} \int d{\bm c} \int d{\bm c}_{1} \int_{0}^{2 \pi} d \varphi \int_{1/2}^{h/\sigma -1/2} dz^{*}_{1}\, \left  (c^{\prime 2}_{=}- c_{=}^{2 } \right)) | {\bm c}_{10} \cdot \widehat{\bm \sigma} | \nonumber \\
&& \times n^{*}(z^{*}_{1}) \Theta \left( -{\bm c}_{10} \cdot \widehat{\bm \sigma} \right)\phi({\bm c}_{1 =},c_{1z}) \phi ({\bm c}_{=},c_{z}).
\end{eqnarray}
 By introducing the center of mass velocity,
\begin{equation}
\label{4.4}
{\bm G} \equiv \frac{ {\bm c}+ {\bm c}_{1}}{2},
\end{equation}
and employing the collision rules, Eqs.\ (\ref{2.6}) and (\ref{2.7}), the above expression can be rewritten as
\begin{eqnarray}
\label{4.5}
\zeta_{=}^{*} & = &- \frac{\gamma+2}{3} \int d{\bm c} \int d{\bm c}_{1} \int_{0}^{2 \pi} d \varphi \int_{1/2}^{h/\sigma -1/2} dz^{*}_{1}\, \left[ \frac{(1+\alpha)^{2}}{4} \left( {\bm c}_{10} \cdot \widehat{\bm \sigma}  \right)^{2} \sigma_{=}^{2} \right.  \nonumber \\
&& \left.  - \frac{1+\alpha}{2}  ({\bm c}_{10} \cdot \widehat{\bm \sigma})(  {\bm c}_{10, =} \cdot {\bm \sigma}_{=})+(1+\alpha) ({\bm c}_{10} \cdot \widehat{\bm \sigma})(  {\bm G}_{=}\cdot {\bm \sigma}_{=})\right] |{\bm c}_{10} \cdot \widehat{\bm \sigma} | \nonumber \\
&& \times n^{*}(z^{*}_{1}) \Theta \left( -{\bm c}_{10} \cdot \widehat{\bm \sigma} \right)\phi({\bm c}_{1 =},c_{1z}) \phi ({\bm c}_{=},c_{z}).
\end{eqnarray}
Notice that ${\bm \sigma}_{=}$  is not a unit vector in the plane $z=\mbox{constant}$, but ${\bm \sigma}_{=}= \widehat{\bm \sigma}_{=}\sin \theta$, with $\widehat{\bm \sigma}_{=}$ being the unit vector. Now we make the change of  variables ${\bm c}_{=}, {\bm c}_{1 =}$ $\rightarrow$ ${\bm c}_{1 =}, {\bm c}_{1}$ and also $\varphi \rightarrow \varphi + \pi$. The latter change is equivalent to ${\bm \sigma}_{=}\rightarrow -{\bm \sigma}_{=}$, while $\widehat{\sigma}_{z}$ remains the same. Then, taking into account that $\phi({\bm c}_{1=}, c_{z}) \phi ({\bm c}_{=}, c_{1z})= \phi({\bm c}_{=}, c_{z}) \phi ({\bm c}_{1 =},c_{1z})$, because of the approximation introduced for the velocity distribution function, Eq\, (\ref{4.1}), it is easily seen that Eq.\ (\ref{4.5}) reduces to
\begin{eqnarray}
\label{4.6}
\zeta_{=}^{*} & =&  \frac{\gamma+2}{3} \int d{\bm c} \int d{\bm c}_{1} \int_{0}^{2 \pi} d \varphi \int_{1/2}^{h/\sigma -1/2} dz^{*}_{1}\, \left[ \frac{1-\alpha^{2}}{4}  | \ {\bm c}_{10} \cdot \widehat{\bm \sigma} |^{3} \right.  \nonumber \\
&& \left.  + \frac{(1+\alpha)^{2}}{4} | {\bm c}_{10} \cdot \widehat{\bm \sigma}|^{3} \widehat{\sigma}_{z}^{2}  +\frac{1+\alpha}{2} |{\bm c}_{10} \cdot  \widehat{\bm \sigma} |^{2}  c_{10z} \widehat{\sigma}_{z}     \right]  \nonumber \\
&& \times n^{*}(z^{*}_{1}) \Theta \left( -{\bm c}_{10} \cdot \widehat{\bm \sigma} \right)\phi({\bm c}_{1 =},c_{1z}) \phi ({\bm c}_{=},c_{z}).
\end{eqnarray}
The exact evaluation of the integrals on the right hand side seems to be a rather complicated task. To derive manageable analytical expressions, some kind of expansion has been considered. Details of the calculations are given in Appendix \ref{b}. Here we only mention the useful relation
\begin{equation}
\label{4.7}
\Theta \left( -{\bm c}_{10 } \cdot \widehat{\bm \sigma} \right) = \Theta \left( - {\bm c}_{10 =} \cdot {\bm \sigma}_{=}\right) - \Theta \left( {\bm c}_{10} \cdot \widehat{\bm \sigma} \right) \Theta \left( - {\bm c}_{10 =} \cdot {\bm \sigma}_{=}\right) + \Theta \left( - {\bm c}_{10} \cdot \widehat{\bm \sigma} \right)  \Theta \left(  {\bm c}_{10 =} \cdot {\bm \sigma}_{=}\right).
\end{equation}
The result is
\begin{eqnarray}
\label{4.8}
\zeta^{*}_{=}& = & \left[ \frac{3 \pi}{2 \left( \gamma +2 \right)} \right]^{1/2} \left(1+\alpha \right)  \nonumber \\
&& \times \left\{ (1-\alpha) \left[ B_{0} \left[ z^{*}|n^{*} \right]  -\frac{3}{2} 
B_{2} \left[ z^{*}|n^{*} \right] \right] + \left[ 1+ \alpha - \frac{(1+3\alpha) \gamma}{2} \right] B_{2} \left[ z^{*}|n^{*} \right] \right\}  \nonumber \\
& & + \mathcal{O} \left( B_{4} \left[z^{*}|n^{*} \right] \right],
\end{eqnarray}
where
\begin{equation}
\label{4.9}
B_{\nu}\left[z^{*}|n^{*} \right] \equiv \int_{1/2}^{h/\sigma-1/2} dz_{1}^{*}\, n^{*}(z_{1}^{*}) \left( z^{*}-z_{1}^{*} \right)^{\nu}.
\end{equation}
In particular, it is $B_{0} \equiv N \sigma^{2} /A$. In a similar way, for the cooling rate of the partial temperature associated to the velocity component perpendicular to the walls one gets
\begin{equation}
\label{4.10}
\zeta_{z}^{*} = \frac{\left(6 \pi\right)^{1/2} \left( 1 +\alpha \right)}{\left( \gamma +2 \right)^{1/2}} \left( 2 - \frac{1+\alpha}{\gamma} \right) B_{2} \left[ z^{*}|n^{*} \right] +
\mathcal{O} \left( B_{4} \left[z^{*}|n^{*}\right] \right).
\end{equation}
Although the above expressions for the partial cooling rates can be written in a more compact way,  Eqs.\ (\ref{4.8}) and (\ref{4.10})   permit to clearly identify the physical meaning of the the several contributions. The term proportional to $1-\alpha$ in the expression of $\zeta_{=}^{*}$ represents the rate of loss of energy associated to the horizontal motion, as a consequence of the horizontal motion itself. Its contribution to the cooling rate $\zeta_{=}$  only depends on the horizontal temperature $T_{=}$ that characterizes the collision frequency. This term vanishes in the elastic limit.  The other term in the expression of $\zeta_{=}^{*}$  is nonzero in general, as it describes the transfer of energy from the horizontal degrees of freedom to the vertical one, which occurs for both elastic and inelastic collisions.  On the other hand, the equation for the rate of variation of the vertical temperature $T_{z}$, does not contain a dissipative contribution  associated to the motion of the particles perpendicular to the walls. This is due to the peculiar geometry of the system. More precisely, this contribution would show up when considering contributions involving higher order functionals $B_{\nu}[z^{*}|n^{*}]$. To the order being considered, all the variations of the kinetic energy in the $z$ direction are due to the interchange with the energy carried out by the motion in the horizontal plane. More about this issue will be said in the final section of the paper. In the limiting case of elastic collisions ($\alpha=1$), it is $\zeta^{*}_{z}T_{z}=-2  \zeta^{*}_{=}T_{=}$, that expresses the kinetic energy conservation in collisions. More precisely the relation reflects that the rate of change of the energy lost (gained) by the vertical degree of freedom is equal to the rate of change  of the energy gained (lost) by the horizontal motion.

Let us notice that, an apparent inconsistency shows up at this point. The assumption that the ICS does exist with the one-particle distribution function having the scaled  form given by Eq.\ (\ref{2.15}), led us to the conclusion that $\gamma$ must be independent of both $z$ and $t$, and also that 
\begin{equation}
\label{4.11}
\zeta^{*}_{=}= \zeta^{*}_{z}, 
\end{equation}
(see Eqs.\ (\ref{2.27}) and (\ref{2.35})). On the other hand, when Eqs. (\ref{4.8}) and (\ref{4.10}) are substituted into the above equation to identify $\gamma$ as a function of $\alpha$, $\sigma$, and $h$, it is trivially seen that $\gamma$ turns out to depend on  $z$ through $B_{2}$, if terms of order $B_{4}$ and beyond are neglected. Nevertheless, this fact does not imply by itself that the assumption on the existence of the ICS is wrong. Expressions in Eqs. (\ref{4.8}) and (\ref{4.10}) have been obtained by introducing the distribution function (\ref{4.1}) and by  expanding in the functionals $B_{\nu}[z^{*}|n^{*}]$, that are not orthogonal with regards to $z^{*}$. Also, the expansion is not directly related with an expansion in powers of $z^{*}$. To overcome this difficulty, the functional $B_{2}[z^{*}|n^{*}]$ that appears in Eqs. (\ref{4.8}) and (\ref{4.10}), will be approximated by 
\begin{equation}
\label{4.12}
B_{2}[z^{*}|n^{*}] \rightarrow \overline{B}_{2} = 
\frac{N \sigma^{4}}{(h-\sigma)^{2} A} \int_{1/2}^ {h/\sigma-1/2} dz^{*}   \int_{1/2}^{ h/\sigma-1/2} dz^{*}_{1} (z^{*}-z^{*}_{1})^{2}
=      \frac{N (h-\sigma)^{2}}{6A}.
\end{equation}
i.e. the local density has been substituted by the average density. It is worth to insist on the idea under this estimate. The expectation is that when all the expansion in the $B_{\nu}$ functionals is  considered, the $z$ dependence on the local density is cancelled out by the power expansion  in $(z^{*}-z^{*}_{1})$. In any case, the accuracy of this approximation, and the existence itself of the ICS must be verified by comparing the derived theoretical predictions with simulation (molecular dynamics) results.

Substitution of Eqs.\ (\ref{4.8}) and (\ref{4.10}) into Eq. (\ref{4.11}), keeping only terms up to $B_{2}$, and employing the approximation given in Eq. (\ref{4.12}), leads to an equation for the temperature ratio $\gamma$ whose physical (positive) solution reads
\begin{equation}
\label{4.13}
\gamma= \frac{d(\alpha)}{e(\alpha)} \left\{ 1+ \sgn\left(d(\alpha)\right) \left[ 1+ \frac{c(\alpha)}{d(\alpha)^{2}} \right]^{1/2} \right\},
\end{equation}
where
\begin{equation}
\label{4.14}
d(\alpha) \equiv 2(1-\alpha)B_{0}+ (5 \alpha-9 ) \overline{B}_{2},
\end{equation}
\begin{equation}
\label{4.15}
e(\alpha) \equiv 2(1+3 \alpha) \overline{B}_{2},
\end{equation}
\begin{equation}
\label{4.16}
c(\alpha) \equiv 16 (1+\alpha) (1+3 \alpha) \overline{B}_{2}^{2},
\end{equation}
and $\sgn(x)$ is the sign function, i.e.
\begin{equation}
\label{4.17}
\sgn(x) = \left\{ \begin{array}{ll}
1 & \mbox{if $x>0 $} \\
0 & \mbox{if $x=0$} \\
-1 & \mbox{if $x<0$} 
\end{array}
\right.
\end{equation}
In the elastic limit, $\alpha=1$, Eq. (\ref{4.13}) reduces to $\gamma=1$, as required by equilibrium statistical mechanics. Also, both cooling rates, $\zeta^{*}_{=}$ and $\zeta_{z}^{*}$, given by Eqs. (\ref{4.8}) and (\ref{4.10}), respectively, vanish, indicating that at equilibrium both (equal) partial temperatures remain constant. Beyond the elastic limit, the approximation given by Eq.\ (\ref{4.13}) is expected to hold for $\epsilon \equiv (h-\sigma)/\sigma \ll 1$, a condition that follows from the expressions of $B_{0}$ and $\overline{B}_{2}$. As already mentioned, this result implies that in the ICS the two partial granular temperatures associated to the vertical and horizontal motions, must be different and, although both decrease monotonically in time, their ratio remains constant.

\section{Molecular Dynamics simulations}
\begin{figure}
\includegraphics[scale=0.4,angle=0]{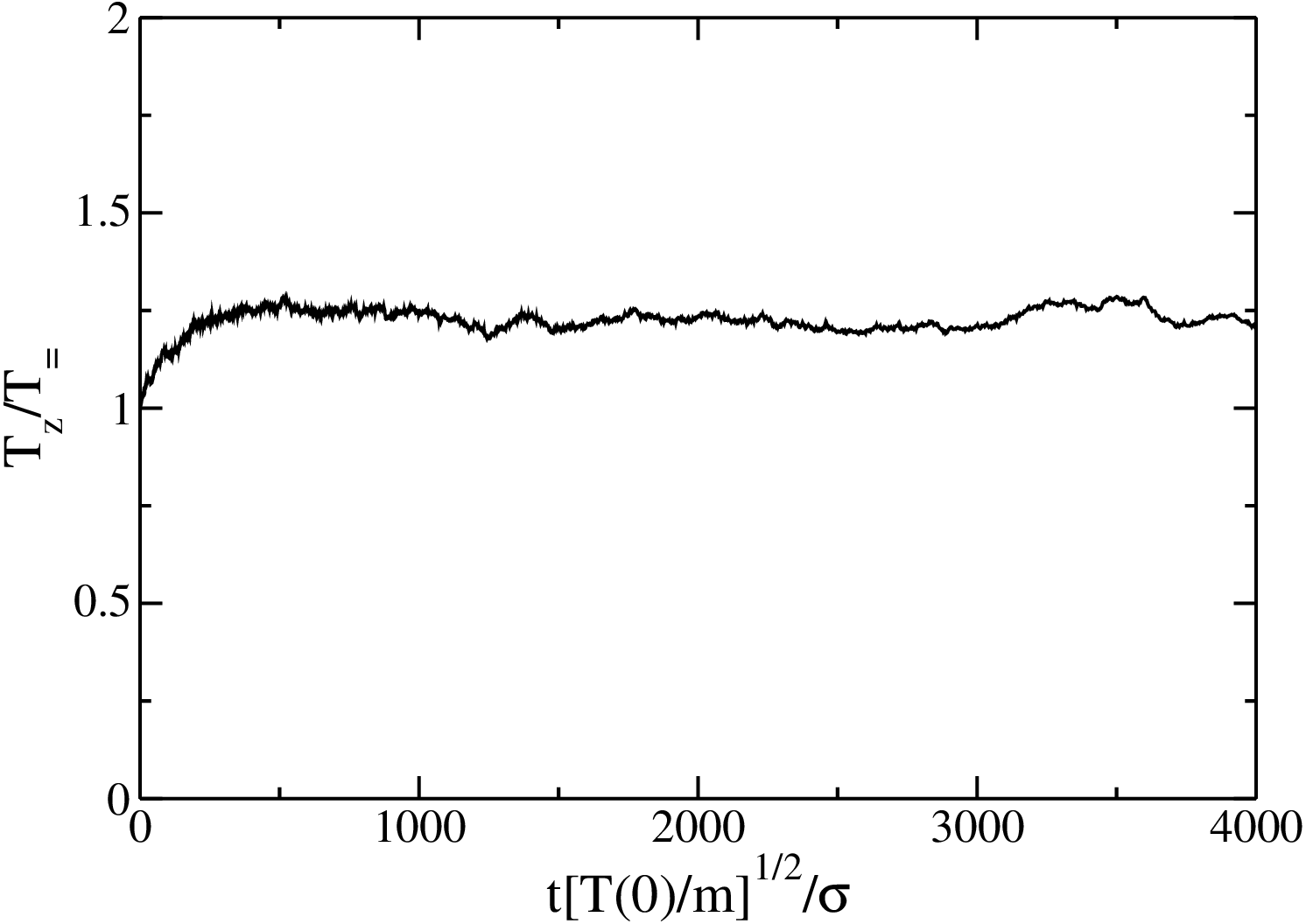}
\caption{Evolution of the temperature ratio $T_{z}(t)/T_{=}(t)$ for a confined system of  $N= 500$ inelastic hard spheres  The average dimensionless density of the system is such that $N \sigma^{2}/A =0.019$, where $N$ is the number of particles, $A$ is the area of each of the confining plates, and $\sigma$ is the diameter of the particles.  The separation of the plates is $h=1.5 \sigma$ and the coefficient of normal restitution is $\alpha=0.95$. Time is measured in the dimensionless units indicated in the label, where $T(0$ is the initial granular temperature. }
\label{fig1}
\end{figure}

In order to investigate the accuracy of the theoretical predictions derived in the previous sections, molecular dynamics (MD) simulations of systems of inelastic hard spheres  have been performed. The two confining plates are squares and periodic boundary conditions have been used in the directions parallel to them. In all cases, it was found that the density profile along the direction perpendicular to the plates and also the temperature ratio, $  T_{z}(t)/ T_{=}(t)$, reached, afeter a transient time period,  steady values, although showing the typical oscilations due to statistical uncertainties. To properly value this behaviour it is worth to hightlight that the constancy of the temperature ratio is a necessary and sufficient condition for the equality of the cooling rates associated to  the two temperature parameters.  The stationary values of the temperature ratio $\gamma$ and the density profile $n(z)$ reported in the following have been averaged over time, once the system is in the ICS. Since the expression for the density profile, Eq. (\ref{3.6}), contains the granular temperature ratio $\gamma $, we report first the results for this latter quantity, whose theoretical prediction is given by Eq. (\ref{4.13}). In the simulations, we never found a dependence of the granular temperatures, $T_{z}$ and $T_{=}$, on the distance to the two plates, consistently with the assumed existence of the ICS.   An example of the way in which the temperature ratio relaxes to its stationary value, is given in Fig. \ref{fig1}. The initial velocity distribution was an isotropic Gaussian with granular temperature $T(0)$.

\begin{figure}
\includegraphics[scale=0.4,angle=0]{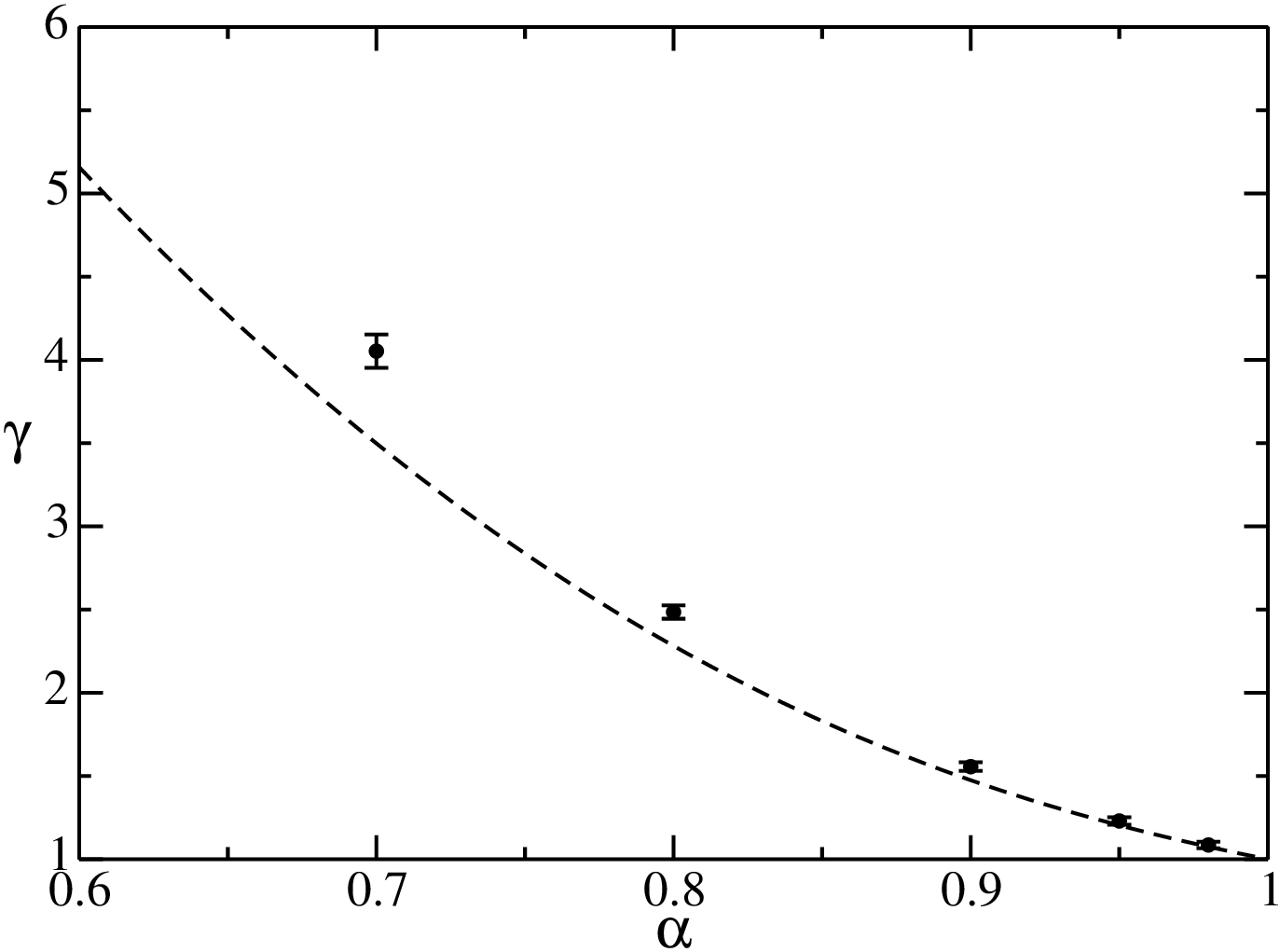}
\caption{Time-independent granular temperature ratio, $\gamma \equiv T_{z}(t)/T_{=}(t)$, for a confined system of inelastic hard spheres in the ICS, as a function  of the coefficient of normal restitution $\alpha$. The symbols are MD simulation results and the dashed line is the theoretical prediction, given by Eq. (\ref{4.13}). The average dimensionless density of the system is such that $N \sigma^{2}/A =0.019$, where $N$ is the number of particles, $A$ is the area of each of the confining plates, and $\sigma$ is the diameter of the particles.  The separation of the plates is $h=1.5 \sigma$. }
\label{fig2}
\end{figure}

Figure \ref{fig2} displays the steady values $\gamma$ of the temperature ratio as a function of the coefficient of normal restitution $\alpha$ for a system of $N=500$ particles. The distance between the two plates is $h=1.5 \sigma$ ($\epsilon=0.5$) and their area $A$ is such that $N \sigma^{2}/A =0.019$. Although  the value of $\epsilon$ is not very small, the agreement between  the theoretical prediction (dashed line) and the simulation results (symbols) is quite satisfactory.  Notice that the value of the granular temperature ratio $\gamma$  changes by a factor of the order of $5$ along the range of values of the coefficient of normal restitution considered, namely $0.6 \leq \alpha <1$).

\begin{figure}
\includegraphics[scale=0.4,angle=0]{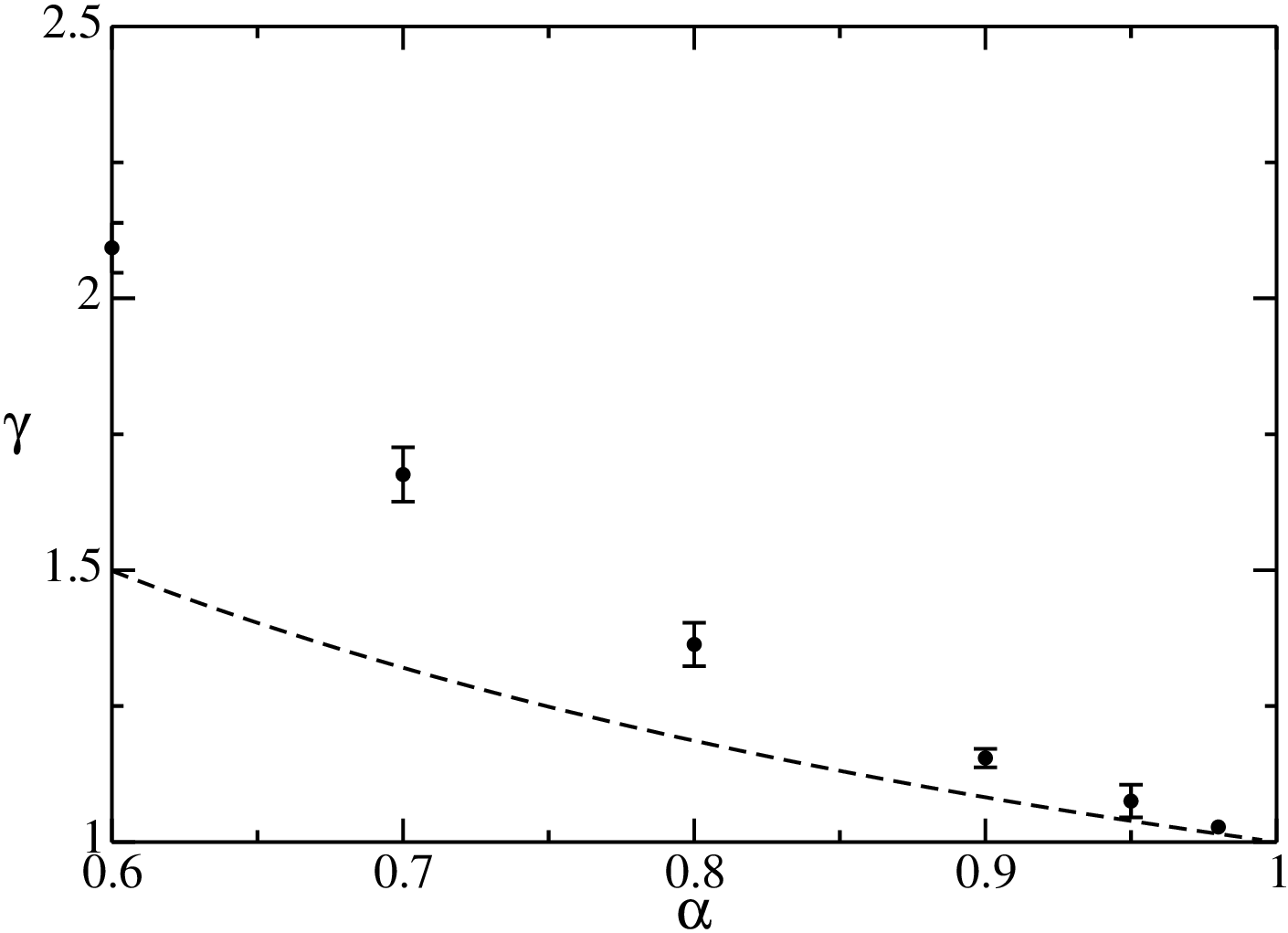}
\caption{The same as in Fig. \ref{fig2}, but now the distance between the two confining plates is $h=1.9 \sigma$. }
\label{fig3}
\end{figure}

Similar results have been found for other values of the parameters defining the state of the system. As expected, the accuracy of the theoretical prediction with the simulation data decreases as the density increases and also as the distance $h$ between the two plates  approaches the limiting value $2\sigma$, beyond which the kinetic equation analyzed here is no longer valid. As an example of this behavior, Fig. \ref{fig3} shows the temperature ratio for the same system as in Fig. \ref{fig2}, but now the separation between the plates is $h=1.9 \sigma$. It is observed that the temperature ratio is now much smaller that in the case reported in Fig. \ref{fig2}, and also that the relative discrepancy  between theory and simulation is larger.

\begin{figure}
\includegraphics[scale=0.4,angle=0]{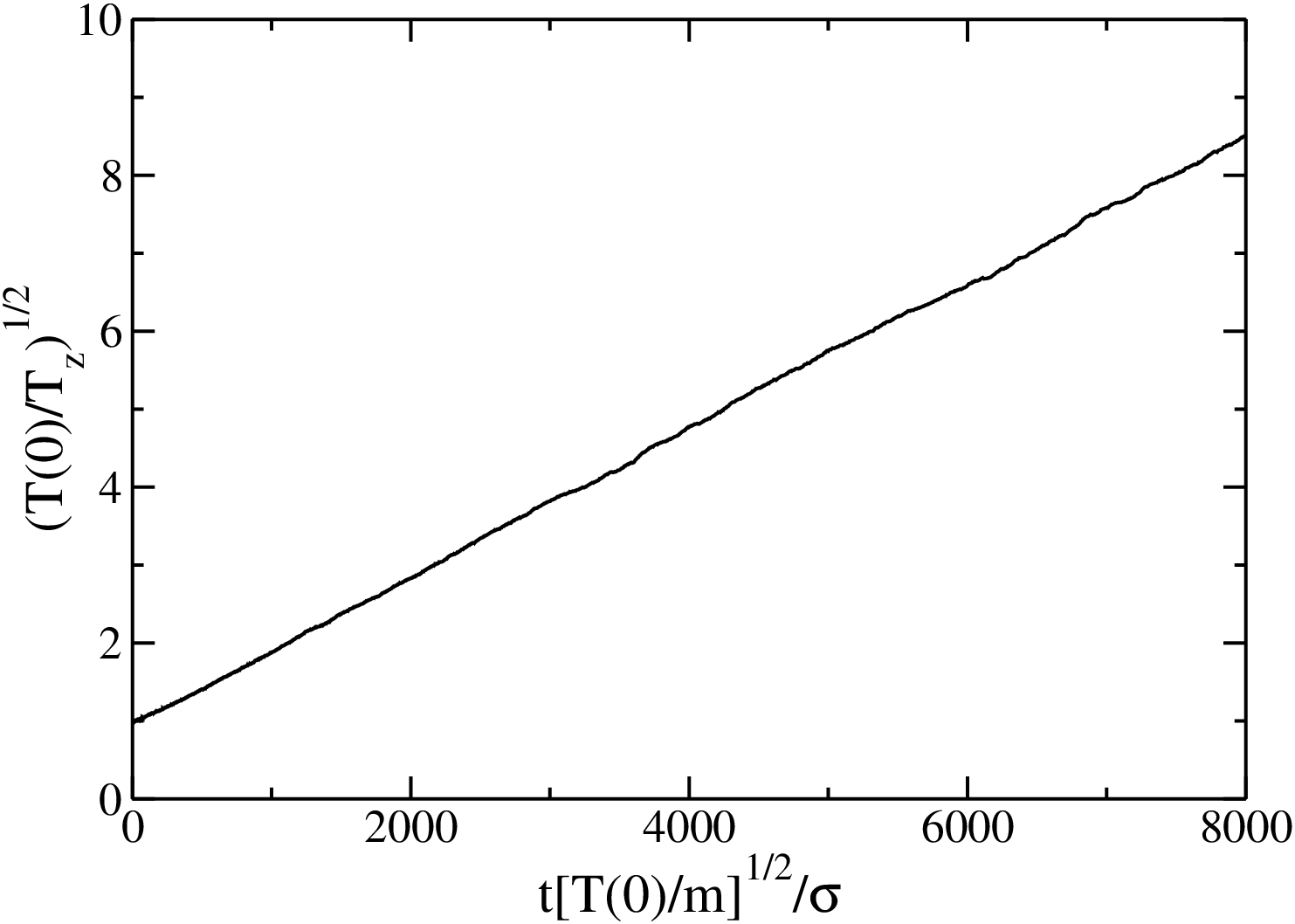}
\caption{MD results for the time evolution of the granular temperature parameter associated to the motion perpendicular to the confining plates. Time is measured in the dimensionless units indicated in the label, where $T(0)$ is the initial granular temperature, $m$ is the mass of the hard spheres, and $\sigma$ their diameter. The number of particles used in the simulation is $= 0500$, the width of the system $h=1.5 \sigma$, and the average density such that $N \sigma^{2}/A=0.019$, with $A$ being the area of each of the plates. }
\label{fig4}
\end{figure}

The accuracy of the predictions for the cooling rates of the partial temperature parameters has also been investigated. From Eqs. (\ref{2.28})-(\ref{2.30} it follows that
\begin{equation}
\label{4.18}
\frac{d T_{z}^{-1/2}}{dt}= \left[ \frac{2(\gamma+2)}{3 \gamma m} \right]^{1/2} \frac{\zeta^{*}_{z}}{\sigma}\, .
\end{equation}
Since the theory predicts that $\zeta^{*}_{z}$ does not depend on time, its value can be measured from the slope of the time evolution of $T_{z}(t)^{-1/2}$, once the temperature ratio is known. This is  illustrated in Fig. \ref{fig4}, where a linear profile is clearly identified. 

\begin{figure}
\includegraphics[scale=0.4,angle=0]{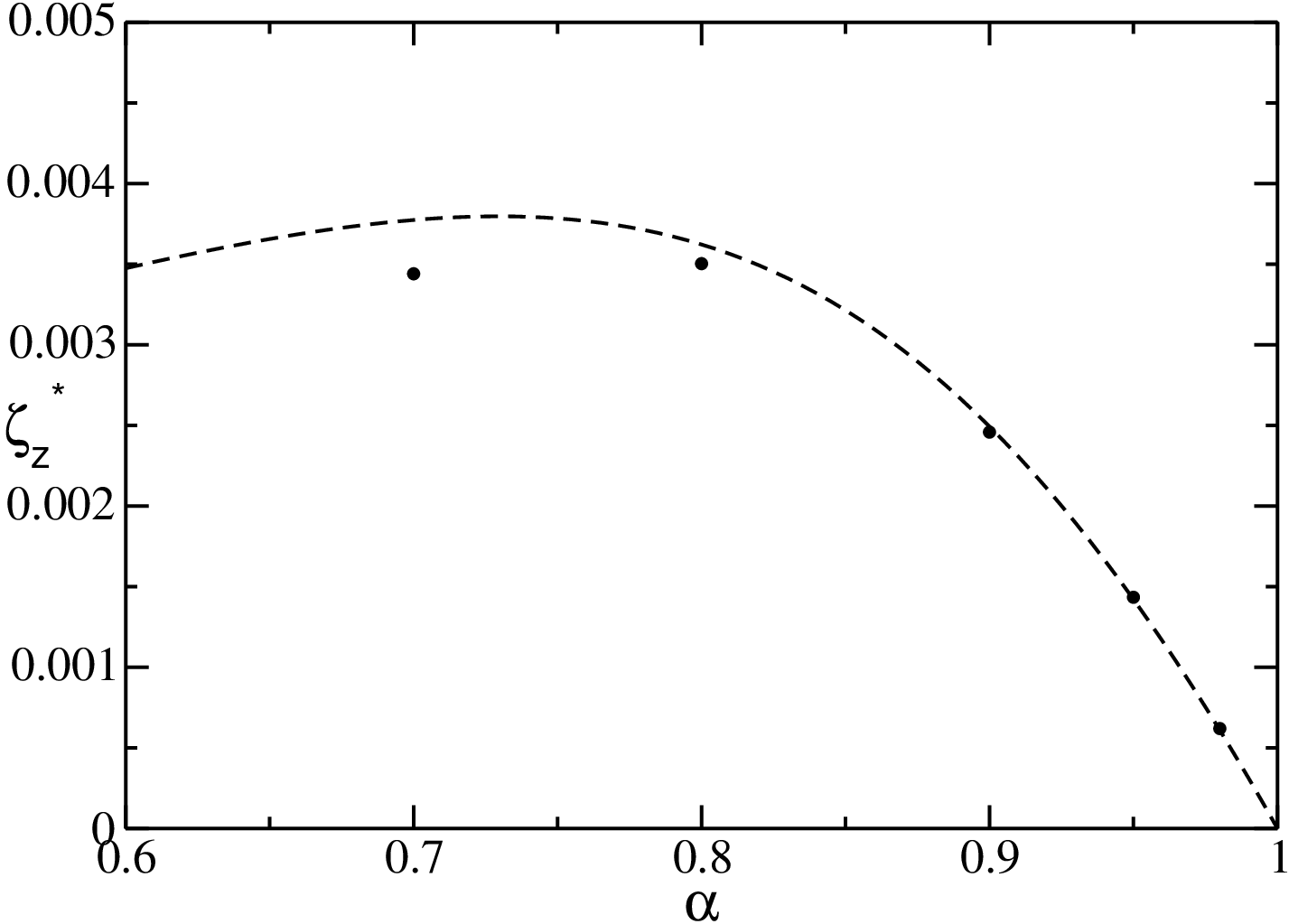}
\caption{The dimensionless cooling rate associated to the motion perpendicular to  the plates, $\zeta_{z}^{*}$,  as a function of the coefficient of normal restitution $\alpha$, for a confined system of hard spheres The number of particles used in the simulation is $N= 0500$, the width of the system $h=1.5 \sigma$, and the average density is such that $N \sigma^{2}/A=0.019$. The symbols are MD simulation results and the dashed line the theoretical prediction given by Eq. (\ref{4.10}).}
\label{fig5}
\end{figure}

The  simulation results for $\zeta^{*}_{z}$ obtained in the above way are compared with the theoretical prediction given in Fig. \ref{fig5}, where the reduced cooling rate is shown as a function of the coefficient of normal restitution $\alpha$. The symbols are the simulation resulta and the dashed line is the theoretical prediction. The other values of the simulation parameters are $N=500$, $h=1.5 \sigma$ and $N \sigma^{2}/A=0.019$. It is seen that the agreement can be considered as quite satisfactory. In particular the theory predicts the non-monotonic behavior  clearly identified in the simulation results.

\begin{figure}
\includegraphics[scale=0.4,angle=0]{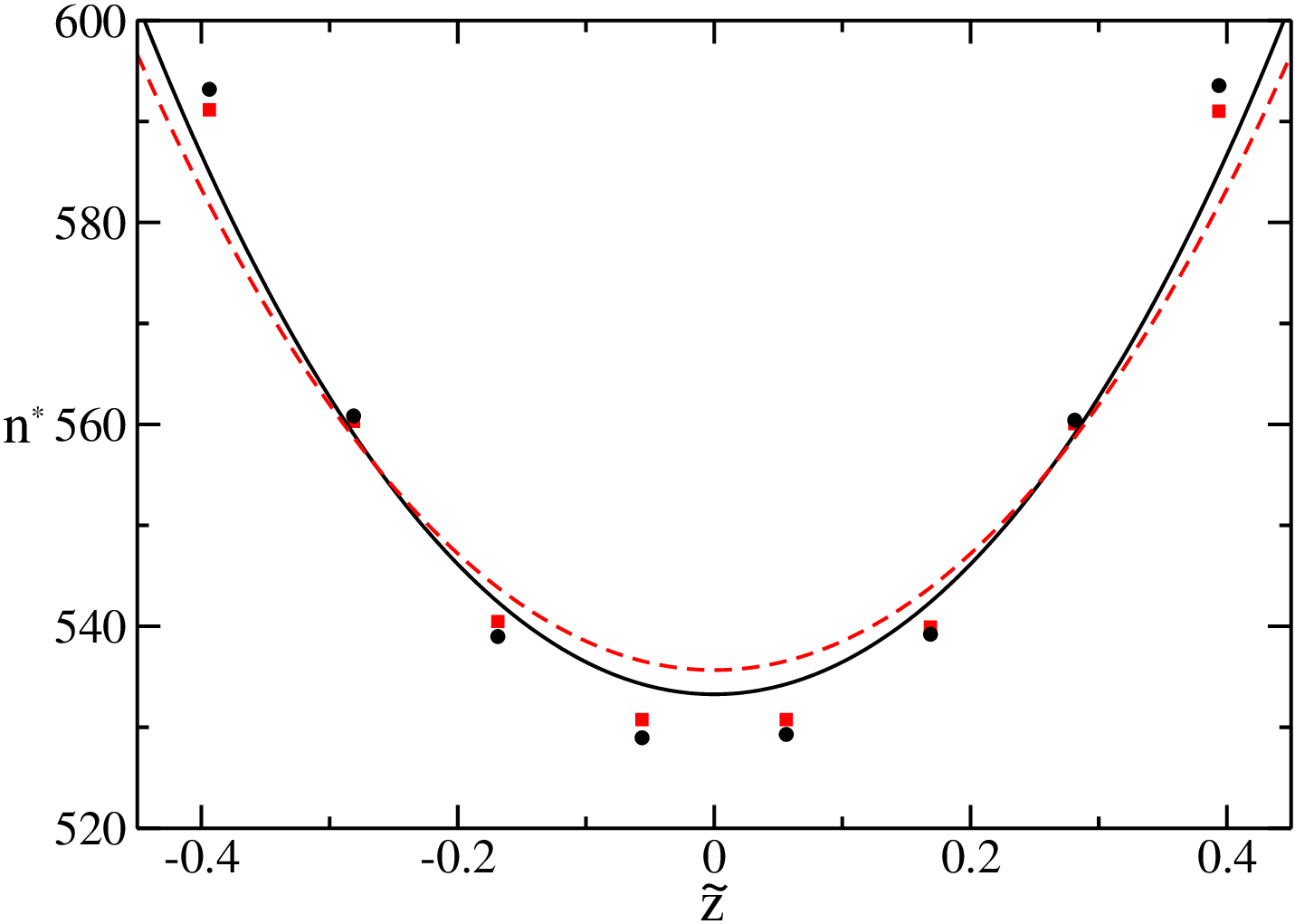}
\caption{(Color online) Dimensionless density profile $n^{*}(z) =n(z) \sigma^{3}$ in the direction perpendicular to the plates in the ICS of a confined quasi-two-dimensional fluid of hard spheres. The solid (black) line is the theoretical prediction in the elastic limit  $\alpha=1$, while the dashed (red) line is for $\alpha=0.9$. The (black) circles   and the (red) squares are simulation data for $\alpha=1$ and $\alpha=0.9$, respectively. The separation betrween the two plates is6$h=1.9 \sigma$, and the density is such that $N/A=0.19 \sigma^{-2}$. }
\label{fig6}
\end{figure}

Next, let us consider the density profile along the direction perpendicular to the plates. Consistently with the theory developed here, it is observed that, after a short transient period, the density profile becomes time-independent, as predicted for the ICS. In Fig. \ref{fig6} the steady density profile is plotted as a function of the scaled distance to the center, $\tilde{z}\equiv \left(z-h/2 \right)/\sigma$ for a system with $h=1.9 \sigma$ and $N\sigma^{2}/A=0.19$. The average density is in this case 10 times larger that in the system considered in Figs. \ref{fig1} and \ref{fig2}. The reason is that, for smaller densities, the curvature of the density profile is rather small and it is hard to identify due to the statistical uncertainties of the simulation data. Also plotted, for reference, is the equilibrium elastic profile ($\alpha=1$). It is seen that, in spite of the relatively high density, there is a good  agreement between theory and simulation. In particular, the effect of inelasticity, although small, can be clearly identified and the perturbation with respect to the elastic case goes in the same direction in both theory and isimulation.  When using Eq.\ (\ref{3.6}) to plot the theoretical prediction for the density profile, the expression derived for the temperature ratio, i.e. the prediction given by Eq.\ (\ref{4.13}), has been employed, so that no parameter has been fitted.

\section{Discussion}
\label{s5}
The aim of this paper is to study the inhomogeneous cooling state (ICS) of a freely evolving granular gas strongly confined  between two parallel hard plates. The marginal distribution functions for the components of the velocity parallel and perpendicular to the plates, have  a scaling form with the time dependence determined by the global local granular temperature of the system, as required for a normal solution and hence for an hydrodynamic description. A consequence of this scaling is that the granular temperature parameters associated to the vertical and horizontal directions are proportional to each other and then the cooling rates of all the granular temperatures are the same. Moreover, this unique cooling rate does not depend on position or time. To put the content of this paper in a proper context, it must be emphasized that the existence of the ICS has not been proven, but assumed. Its justification lies on the comparison of the predicted properties of the state with numerical simulations. In any case, the accuracy of the prediction for the partial cooling rates and the steady temperature ratio discussed here, provide a strong evidence of the existence of the ICS.

The eventual tendency of the freely evolving granular gas towards a state (the ICS) in which the ratio of the granular temperatures is constant, manifests itself in a quite particular behavior of the partial temperatures. Consider a confined granular gas that is not in the ICS, and assume that the one-particle distribution function of the gas can be accurately approximated  by a gaussian with two different temperatures as given in Eq. (\ref{4.1}). This is expected to be true close to the ICS. Then, the cooling rates are estimated by Eqs. (\ref{4.8}) an (\ref{4.10}), respectively. Keeping only the lowest order approximation, i.e. up to $B_{2}$, it is seen that the cooling rate associated to the motion perpendicular to the plates,  $\zeta_{z}^{*}$, is positive when $\gamma > (1+\alpha)/2$ or, equivalently,
\begin{equation}
\label{5.1}                                                                                                                                                                         
T_{z}> \frac{1+\alpha}{2}\, T_{=}.
\end{equation}
Taking into account that $0< \alpha \leq 1$, it is seen that the above condition is verified if $T_{z} >T_{=}$, but also if
\begin{equation}
\label{5.2}
T_{=}> T_{z} >\frac{1+\alpha}{2}\, T_{=}.
\end{equation}
A positive $\zeta_{z}$ means that the temperature $T_{z}$ is decreasing in time. Therefore, it follows that when condition (\ref{5.2}) is fulfilled, energy is being transferred from the vertical degree of freedom to the horizontal ones, although the partial granular temperature of the former is lower than the temperature of the later. On the other hand, $\zeta_{z}^{*}$ is negative if
\begin{equation}
\label{5.3}
T_{z} < \frac{1+\alpha}{2}\, T_{=}< T_{=},
\end{equation}
and in this  case energy goes from the degrees of freedom with a larger temperature to the degree with a smaller one. i.e. the usual behavior. Let us now analyze the behavior of the  temperature parameter associated to the horizontal motion, $T_{=}$.  As discussed below Eqs. (\ref{4.8})-(\ref{4.10}), the cooling rate describing the energy interchanged by the horizontal motion with the vertical one, up to order $B_{2}$ is
\begin{equation}
\label{5.4}
\zeta_{=}^{*(z)} = \left[ \frac{3 \pi}{2 (\gamma +2)} \right]^{1/2} (1+\alpha) \left[ (1+\alpha)- \frac{(1+3 \alpha) \gamma}{2} \right] B_{2} \left[ z^{*}|n^{*}\right].
\end{equation}
Then, $\zeta_{=}^{*(z)} >0$ is equivalent to
\begin{equation}
\label{5.5}
T_{z} < T_{=}< \frac{2(1+\alpha)}{1+3 \alpha}\, T_{=}.
\end{equation}
This can be accomplished both if $T_{z} <T_{=}$ or 
\begin{equation}
\label{5.6}
\frac{2(1+\alpha)}{1+3 \alpha}\, T_{=}>T_{z} > T_{=}.
\end{equation}
In the last case, energy is again being supplied by the degrees of freedom with a lower granular temperature parameter to the degree with a higher temperature.
 Proceeding in the same way, it is seen that when the temperature associated to the motion in the plane increases, the temperature associated to this motion is smaller than the vertical temperature, i.e. a ``normal'' behavior.  The above discussion is based on the approximated expressions for the cooling rates given in Eqs. (\ref{4.8}) and (\ref{4.10}), which are expected to provide an accurate description of the time evolution of the temperature parameters o the towards the ICS, in the system described by the Boltzmann equation (\ref{2.1}).  
  
 It is worth to emphasize that the apparent anomalous behavior predicted in some cases for the direction of the energy flux between different degrees of freedom, has been described in terms of granular temperatures, which by no means are equivalent or even similar to the temperature concept used in thermodynamics, but they are just a measure of the local kinetic energy of the macroscopic hard spheres. Consequently,  that behavior can not be understood as violating any macroscopic fundamental law for molecular systems and, in particular, the key property of the thermodynamic temperature as defined by Clausius that heat can never spontaneously flows from cold to hot \cite{FyW15}
 
 The analysis of the ICS presented here is the essential first step needed for the derivation of hydrodynamic equations for strongly confined systems of inelastic hard spheres. Only after this macroscopic description has been worked out, it will be  possible to answer theoretically many questions related with the rich phenomenology exhibited by the system and mentioned in the Introduction. For instance, it has been shown recently that an effective hydrodynamic model that is consistent with the ideas reported here in the quasielastic limit, predicts  the existence of an instability leading to the formation of a density cluster \cite{MGyB19a}. This result can be related with some experimental findings in which the coexistence between a liquid-like phase and a gas-like phase has been reported in a system similar to the one considered in this paper, although with a larger separation between the two plates \cite{RCBHyS11,CRBHyS12}. It is worth to mention that the phenomenon of the coexistence of two phases in a narrow vibrated granular gas has been rather elusive, in the sense that several models proposed in the literature were unable, in our opinion, to provide a fully satisfactory explanation of it \cite{KyT15,GMyT13,BRyS13,BBGyM16}.

\acknowledgments

This research was supported by the Ministerio de Econom\'{\i}a, Industria  y Competitividad  (Spain) through Grant No. FIS2017-87117-P (partially financed by FEDER funds).

\appendix

\section{Equation for the density profile}
\label{a}

Here the calculations leading to integro-differential equation for the density profile, Eq. (\ref{3.1}), will be outlined. Multiplication of Eq.\ (\ref{2.37}) by $c_{i}$ and afterwards integration over ${\bm c}$ yields
\begin{equation}
\label{a.1}
\delta_{i,z}\,  \frac{3 \gamma}{2(\gamma+2)} \frac{\partial \ln n^{*}(z^{*})}{\partial z^{*}} = \int d{\bm c}\,  c_{i} J^{*}[z^{*},{\bm c}|\phi],
\end{equation}
\begin{eqnarray}
\label{a.2}
\int d{\bm c}\,  c_{i} J^{*}[z^{*},{\bm c}|\phi ]
& = &\int d{\bm c} \int d{\bm c}_{1} \int_{0}^{2\pi} d \varphi \int_{1/2}^{h/\sigma-1/2} d z^{*}_{1}\, (c^{\prime}_{i}-c_{i} ) | {\bm c}_{10} \cdot \widehat{\bm \sigma} | n^{*}(z^{*}_{1}) \nonumber \\
&& \times \Theta \left( -{\bm c}_{10} \cdot \widehat{\bm \sigma} \right)\phi({\bm c}_{1 =},c_{1z}) \phi ({\bm c}_{=},c_{z}),
\end{eqnarray}
where the dimensionless form of the property given in Eq.\ (\ref{2.5}) and the definition of the temperature ratio $\gamma$, have been used. Interchanging ${\bm c}$ and ${\bm c}_{1}$, and taking into account momentum conservation in collisions, it is easily seen that 
\begin{eqnarray}
\label{a.3}
 \int d{\bm c}\ c_{i} J^{*}[z^{*},{\bm c}|\phi] & = & -\frac{1}{2} \int d{\bm c} \int d{\bm c}_{1} \int_{0}^{2\pi} d \varphi \int_{1/2}^{h/\sigma-1/2} d z^{*}_{1}\, (c^{\prime}_{i}-c_{i} )  ({\bm c}_{10} \cdot \widehat{\bm \sigma})  n^{*}(z^{*}_{1}) \nonumber \\
&& \times \phi({\bm c}_{1 =},c_{1z}) \phi ({\bm c}_{=},c_{z}), \nonumber \\
&=& - \frac{1+\alpha}{4 } \int d{\bm c} \int d{\bm c}_{1} \int_{0}^{2\pi} d \varphi \int_{1/2}^{h/\sigma-1/2} d z^{*}_{1}\, ({\bm c}_{10} \cdot \widehat{\bm \sigma})^{2}  \widehat{\sigma}_{i}  n^{*} (z^{*}_{1}) \nonumber \\
&& \times  \phi({\bm c}_{1 =},c_{1z}) \phi ({\bm c}_{=},c_{z}) \nonumber \\
&=&  - \frac{1+\alpha}{4 } \int d{\bm c} \int d{\bm c}_{1} \int_{0}^{2\pi} d \varphi \int_{1/2}^{h/\sigma-1/2} d z^{*}_{1}\, \left[ \sum_{j} \left(c_{1j}^{2}+c_{j}^{2}  \right)\widehat{\sigma}_{j}^{2} \right]  \widehat{\sigma}_{i} \nonumber \\
&&\times n^{*}(z^{*}_{1})\phi({\bm c}_{1 =},c_{1z}) \phi ({\bm c}_{=},c_{z}).
\end{eqnarray}
The velocity integrals can be expressed as functions of the temperature ratio  using Eqs.\ (\ref{2.25})-(\ref{2.28}). The result is
\begin{equation}
\label{a.4}
\int d{\bm c}\, c_{i} J^{*}[z^{*},{\bm c}|\phi] = - \frac{3(1+\alpha)}{4 (\gamma +2)} \int_{0}^{2 \pi} d \varphi \int_{1/2}^{h/\sigma-1/2} d z^{*}_{1}\,  \left( \sigma_{=}^{2} + \gamma \widehat{\sigma}_{z}^{2} \right) n^{*}(z^{*}_{1}) \widehat{\sigma}_{i},
\end{equation}
with $\sigma_{=}^{2} = 1- \widehat{\sigma}_{z}^{2}$. Next the expression of $\widehat{\bm \sigma}$ in terms of the angles $\varphi$ and $\theta$ is introduced and  the integral over $\varphi$ is carried out to get
\begin{equation}
\label{a.5}
\int d{\bm c}\, c_{i} J^{*}[z^{*},{\bm c}|\phi] = -\delta_{i,z} \frac{3 \pi (1+\alpha)}{2 (2+\gamma)}  \int_{1/2}^{h/\sigma-1/2} d z^{*}_{1}\, \left[ \cos \theta+ (\gamma-1) \cos^{3} \theta \right] n^{*} (z_{1}^{*}).
\end{equation}
Now, the relation $\cos\theta = z^{*}_{1}-z^{*}$ is employed, and Eq. (\ref{3.1}) is easily obtained from Eqs. (\ref{a.1}) and (\ref{a.5}).

\section{Calculation of the cooling rate for the horizontal temperature $T_{=}$}
\label{b}
The derivation of Eq. (\ref{4.8}) from Eq. (\ref{4.6}) will be outlined in this appendix. Each of the terms inside the square brackets on the right-hand-side of Eq. (\ref{4.6}) will be analyzed separately. Consider first
\begin{eqnarray}
\label{b.1}
\zeta_{=}^{*(1)}& \equiv& \frac{\left( \gamma +2 \right) \left(1-\alpha^{2} \right)}{12}  \int d{\bm c} \int d{\bm c}_{1} \int_{0}^{2\pi} d \varphi \int_{1/2}^{h/\sigma -1/2} d z^{*}_{1}\, |  {\bm c}_{10} \cdot \widehat{\bm \sigma} |^{3}  n^{*}(z^{*}_{1}) \nonumber \\
 && \times \Theta \left( -{\bm c}_{10} \cdot \widehat{\bm \sigma} \right)\phi({\bm c}_{1 =},c_{1z}) \phi ({\bm c}_{=},c_{z}).
\end{eqnarray}
Now, the identity (\ref{4.7}) is employed, taking into account that
\begin{eqnarray}
\label{b.2}
 \int d{\bm c} \int d{\bm c}_{1}  \left( {\bm c}_{10} \cdot  \widehat{\bm \sigma}  \right)^{3} \Theta \left( {\bm c}_{10} \cdot \widehat{\bm \sigma} \right) \Theta \left( -{\bm c}_{10 =} \cdot {\bm \sigma}_{=}\right) \phi({\bm c}_{1 =},c_{1z}) \phi ({\bm c}_{=},c_{z}) \nonumber \\
 =
 -\int d{\bm c} \int d{\bm c}_{1}   \left( {\bm c}_{10} \cdot  \widehat{\bm \sigma}  \right)^{3} \Theta \left( - {\bm c}_{10} \cdot \widehat{\bm \sigma} \right) \Theta \left( {\bm c}_{10 =} \cdot {\bm \sigma}_{=}\right) \phi({\bm c}_{1 =},c_{1z}) \phi ({\bm c}_{=},c_{z}).
\end{eqnarray} 
 This allows to rewrite Eq. (\ref{b.1}) as
 \begin{eqnarray}
 \label{b.3}
\zeta_{=}^{*(1)}& = & - \frac{\left( \gamma +2 \right) \left(1-\alpha^{2} \right)}{12}  \int d{\bm c} \int d{\bm c}_{1} \int_{0}^{2\pi} d \varphi \int_{1/2}^{h/\sigma -1/2} d z^{*}_{1}\,   \left( {\bm c}_{10} \cdot  \widehat{\bm \sigma}  \right)^{3}   n^{*}(z^{*}_{1}) \nonumber \\
 && \times \left[ \Theta \left( - {\bm c}_{10=} \cdot {\bm \sigma} \right)- 2 \Theta \left(-  {\bm c}_{10 =} \cdot {\bm \sigma}_{=}\right) \Theta \left(   {\bm c}_{10} \cdot  \widehat{\bm \sigma}\right) \right] \phi({\bm c}_{1 =},c_{1z}) \phi ({\bm c}_{=},c_{z}).
\end{eqnarray} 
 Next, the expansion
 \begin{equation}
 \label{b.4}
 \left( \widehat{\bm c}_{10} \cdot {\bm \sigma}  \right)^{3}= \left(  {\bm c}_{10 =} \cdot  {\bm \sigma}_{=}\right)^{3} + 3 \left(  {\bm c}_{10 =} \cdot  {\bm \sigma}_{=}\right)^{2} c_{10z}  \widehat{\sigma}_{z}+ 3   {\bm c}_{10 =} \cdot  {\bm \sigma}_{=} \left( c_{10z}  \widehat{\sigma}_{z} \right)^{2}+ \left ( c_{10z} \widehat{\sigma}_{z} \right)^{3}
 \end{equation}
 is introduced into the right hand side of the above equation. An straightforward calculation gives
 \begin{eqnarray}
 \label{b.5}
 \int d{\bm c} \int d{\bm c}_{1} \int_{0}^{2\pi} d &\varphi& \int_{1/2}^{h/\sigma -1/2} d z^{*}_{1}\, \left(  {\bm c}_{10 } \cdot \widehat{\bm \sigma}\right)^{3}  n^{*}(z^{*}_{1})  \Theta \left( -{\bm c}_{10=} \cdot {\bm \sigma}_{=}\right) \phi({\bm c}_{1 =},c_{1z}) \phi ({\bm c}_{=},c_{z})    \nonumber \\
&=& -2 (2\pi)^{1/2} \left( \frac{3}{\gamma+2} \right)^{3/2} \int_{1/2}^{h/\sigma-1/2} dz^{*}_{1} \, n^{*}(z^{*}_{1}) \sin^{3}  \theta  \nonumber \\
& & -18 \left( \frac{3\pi}{2} \right)^{1/2} \frac{\gamma}{(\gamma +2)^{3/2}} \int_{1/2}^{h/\sigma-1/2} dz^{*}_{1} \, n^{*}(z^{*}_{1}) \sin \theta \cos^{2} \theta.
\nonumber \\
 \end{eqnarray}
The term on the right-hand-side of Eq. (\ref{b.3}) containing the product of two Heaviside functions can not be evaluated analytically, at least in a simple way. Then, we  formally expand,
\begin{eqnarray}
\label{b.6}
\Theta( {\bm c}_{10} \cdot \widehat{\bm \sigma} )& = &\Theta \left(  {\bm c}_{10 =} \cdot  {\bm \sigma}_{=}+ c_{10z}\widehat{\sigma}_{z} \right) = \Theta ({\bm c}_{10=} \cdot {\bm \sigma}_{=} )+ \delta \left(  {\bm c}_{10=} \cdot {\bm \sigma}_{=}\right) c_{10z} \widehat{\sigma}_{z} \nonumber \\
&& + \frac{1}{2} \delta^{\prime} \left( {\bm c}_{10 =} \cdot  {\bm \sigma}_{=}\right) c_{10z}^{2} \widehat{\sigma}^{2}_{z}  + \cdots,
\end{eqnarray}
where $\delta^{\prime}$ is the derivative of the Dirac delta function. Then, it is easily seen that
\begin{eqnarray}
\label{b.7}
 \int d{\bm c} \int d{\bm c}_{1} \int_{0}^{2\pi} d & \varphi& \int_{1/2}^{h/\sigma -1/2} d z^{*}_{1}\, \left(  {\bm c}_{10 } \cdot  {\bm \sigma} \right)^{3}n^{*}(z^{*}_{1}) \Theta \left( - {\bm c}_{10 =}\cdot  {\bm \sigma}_{=}\right) \Theta \left(  {\bm c}_{10} \cdot \widehat{\bm \sigma} \right)  \nonumber \\
 && \times \phi({\bm c}_{1 =},c_{1z}) \phi ({\bm c}_{=},c_{z}) = \mathcal{O} \left(B_{4}[z^{*}|n^{*}] \right).
\end{eqnarray}
Here the relation $\widehat{\sigma}_{z} = \cos  \theta$ has been employed and $B_{\nu}[z^{*}|n^{*}]$ is the functional defined in Eq.\ (\ref{4.9}). By consistency, in the first term on the right hand side of Eq. (\ref{b.5})  the expansion $\sin^{3} \theta = 1-3( \cos^{2} \theta)/2 + \mathcal{O} (\cos^{4}\theta)$ is considered and, similarly,  in the second term the relation $\sin \theta =1+\mathcal{O} (\cos^{2} \theta)$ is used. In this way one gets
\begin{equation}
\label{b.8}
\zeta_{=}^{*(1)}=  \left[ \frac{3 \pi}{2 (\gamma +2)} \right]^{1/2}  (1-\alpha^{2}) \left[  B_{0}+ \frac{3}{2} \left( \gamma-1) \right) B_{2}[z^{*}|n^{*}) \right]
+ \mathcal{O} \left( B_{4}[z^{*}|n^{*} ] \right).
\end{equation}
The next contribution to $\zeta^{*}_{=}$ as given in Eq. (\ref{4.6}) to be analyzed is
\begin{eqnarray}
\label{b.9}
\zeta_{=}^{*(2)} &=&  -\frac{(\gamma +2)(1+\alpha)^{2}}{12}  \int d{\bm c} \int d{\bm c}_{1} \int_{0}^{2\pi} d \varphi \int_{1/2}^{h/\sigma -1/2} d z^{*}_{1}\, \left(    {\bm c}_{10}  \cdot \widehat{\bm \sigma}\right)^{3} \widehat{\sigma} _{z}^{2} n^{*}(z^{*}_{1}) \nonumber \\ 
&& \times \Theta \left( -  {\bm c}_{10}\cdot  \widehat{\bm \sigma}\right) \phi({\bm c}_{1 =},c_{1z}) \phi ({\bm c}_{=},c_{z}) \nonumber \\
&=& -\frac{(\gamma +2)(1+\alpha)^{2}}{12}  \int d{\bm c} \int d{\bm c}_{1} \int_{0}^{2\pi} d \varphi \int_{1/2}^{h/\sigma -1/2} d z^{*}_{1}\, \left(   {\bm c}_{10 =} \cdot {\bm \sigma}_{=}\right)^{3} \widehat{\sigma}_{z}^{2} n^{*}(z^{*}_{1})  \nonumber \\ 
&& \times \Theta \left( - {\bm c}_{01 =} \cdot  {\bm \sigma}_{=} \right) \phi({\bm c}_{1 =},c_{1z}) \phi ({\bm c}_{=},c_{z}) + \mathcal{O} \left( B_{4}[z^{*}|n^{*}] \right).
\end{eqnarray}
In the last transformation use has been made of Eq. (\ref{4.7}). It is now a simple task to show that
\begin{equation}
\label{b.10}
\zeta_{=}^{*(2)} = \left[ \frac{3 \pi}{2( \gamma+2)} \right]^{1/2}\, (1+\alpha)^{2} B_{2} [z^{*}|n^{*}] +\mathcal{O} \left( B_{4}[z^{*}|n^{*}] \right).
\end{equation}
The last contribution to $\zeta^{*}_{=}$ in Eq.\ (\ref{4.6}) is analyzed in a similar way,
\begin{eqnarray}
\label{b.11}
\zeta_{=}^{*(3)} &=& \frac{(\gamma+2)(1+\alpha)}{6}  \int d{\bm c} \int d{\bm c}_{1} \int_{0}^{2\pi} d \varphi \int_{1/2}^{h/\sigma -1/2} d z^{*}_{1}\, \left(  {\bm c}_{10} \cdot \widehat{\bm \sigma}\right)^{2}     c_{10z}  \widehat{\sigma}_{z} n^{*}(z_{1}^{*}) \nonumber \\
&& \times  \Theta \left( -  {\bm c}_{10} \cdot \widehat{\sigma} \right) \phi({\bm c}_{1 =},c_{1z}) \phi ({\bm c}_{=},c_{z}) \nonumber \\
&=& - \left( \frac{6 \pi}{\gamma +2} \right)^{1/2} \gamma (1+\alpha) B_{2}[z^{*}|n^{*}]  +\mathcal{O} \left( B_{4}[z^{*}|n^{*}] \right).
\end{eqnarray}
By putting together all the above results and computing $\zeta^{*}_{=}=\zeta^{*(1)}_{=}+ \zeta^{*(2)}_{=}+\zeta^{*(3)}_{=}$, Eq. (\ref{4.8}) follows.

\section{Calculation of the cooling rate for the vertical temperature $T_{z}$}
\label{c}
The calculation of $\zeta_{z}$ is similar to thet of $\zeta_{=}$ discussed in the previous section. From Eq. (\ref{2.31}), using the property (\ref{2.5}) and the collision rules (\ref{2.6}) and (\ref{2.7}),
\begin{eqnarray}
\label{c.1}
\zeta_{z}^{*} & = &  - \frac{2(\gamma+2)}{3 \gamma} \int d{\bm c} \int d{\bm c}_{1} \int_{0}^{2\pi} d \varphi \int_{1/2}^{h/\sigma -1/2} d z^{*}_{1}\, \left( c_{z}^{ \prime 2}-c_{z}^{2} \right)| {\bm c}_{10} \cdot \widehat{\bm \sigma} | \nonumber \\
&& \times n^{*}(z^{*}_{1}) \Theta \left( -{\bm c}_{10} \cdot \widehat{\bm \sigma} \right)\phi({\bm c}_{1 =},c_{1z}) \phi ({\bm c}_{=},c_{z}) \nonumber \\
&=& - \frac{2(\gamma+2)}{3 \gamma} \int d{\bm c} \int d{\bm c}_{1} \int_{0}^{2\pi} d \varphi \int_{1/2}^{h/\sigma -1/2} d z^{*}_{1}\,  \nonumber \\
&& \times  \left[  \frac{(1+\alpha)^{2}}{4}\, \left( {\bm c}_{10} \cdot \widehat{\bm \sigma} \right)^{2} \widehat{\sigma}_{z}^{2} - \frac{1+\alpha}{2} {\bm c}_{10} \cdot \widehat{\bm \sigma} c_{10z} \widehat{\sigma}_{z}  +(1+\alpha) {\bm c}_{10} \cdot \widehat{\bm \sigma} G_{z} \widehat{\sigma}_{z} \right] \nonumber \\
&& \times | {\bm c}_{10} \cdot \widehat{\bm \sigma} |  n^{*}(z^{*}_{1}) \Theta \left( -{\bm c}_{10} \cdot \widehat{\bm \sigma} \right)\phi({\bm c}_{1 =},c_{1z}) \phi ({\bm c}_{=},c_{z}).
\end{eqnarray}
This is written in the abbreviated form
\begin{equation}
\label{c.2}
\zeta^{*}_{z}= \zeta^{*(1)}_{z}+ \zeta^{*(2)}_{z} +  \zeta^{*(3)}_{z},
\end{equation}
where each of the three terms corresponds to each of the three addends inside the square brackets on the right hand side of Eq. ({\ref{c.1}), and so they can be easily identified by comparison. Below each of the three terms is evaluated separately.
\begin{eqnarray}
\label{c.3}
\zeta_{z}^{*(1)} & = &\frac{(\gamma + 2 )(1+\alpha)^{2}}{6 \gamma} \int d{\bm c} \int d{\bm c}_{1} \int_{0}^{2\pi} d \varphi \int_{1/2}^{h/\sigma -1/2} d z^{*}_{1}\, \left( {\bm c}_{10 =} \cdot {\bm \sigma}_{=}\right)^{3} \widehat{\sigma}_{z}^{2} \nonumber \\
&& \times n^{*}(z_{1}^{*}) \Theta \left( - {\bm c}_{10 =} \cdot {\bm \sigma}_{=}\right) \phi({\bm c}_{1 =},c_{1z}) \phi ({\bm c}_{=},c_{z}) +  \mathcal{O} \left( B_{4}[z^{*}|n^{*} ] \right).
\end{eqnarray}
The integrations over the velocities and the azimuthal angle are easily evaluated with the result
\begin{equation}
\label{c.4}
\zeta_{z}^{*(1)} = - \left( \frac{ 6 \pi}{\gamma +`2} \right)^{1/2} \frac{(1+\alpha)^{2}}{\gamma} B_{2} [z^{*}|n^{*}] + \mathcal{O} \left( B_{4}[z^{*}|n^{*} ] \right).
\end{equation}
In a similar way,
\begin{eqnarray}
\label{c.5}
\zeta_{z}^{*(2)} &= & \frac{4 (\gamma +2) (1+\alpha)}{3 \gamma} \int d{\bm c} \int d{\bm c}_{1} \int_{0}^{2\pi} d \varphi \int_{1/2}^{h/\sigma -1/2} d z^{*}_{1}\, c_{10=} c_{10z}^{2} \widehat{\sigma}_{z}^{2} n^{*}(z^{*}_{1}) \nonumber \\
&& \times \phi({\bm c}_{1 =},c_{1z}) \phi ({\bm c}_{=},c_{z}) +  \mathcal{O} \left( B_{4}[z^{*}|n^{*} ] \right) \nonumber \\
&=& 2 \left( \frac{6 \pi}{\gamma +2} \right)^{1/2} (1+\alpha) B_{2} [z^{*}|n^{*}] + \mathcal{O} \left( B_{4}[z^{*}|n^{*} ] \right).
\end{eqnarray}
Finally, it is
\begin{equation}
\label{c.6}
\zeta^{*(3)}_{z} =  \mathcal{O} \left( B_{4}[z^{*}|n^{*} ] \right).
\end{equation}
Use of Eqs. (\ref{c.4}), (\ref{c.5}), and (\ref{c.6}) leads to Eq. (\ref{4.10}).

\end{document}